\pdfoutput=1

\documentclass{article}
\usepackage{tabularx}
\usepackage{longtable}
\usepackage{comment}
\usepackage[preprint]{neurips_2023}
\usepackage[utf8]{inputenc} 
\usepackage[T1]{fontenc}    
\usepackage{hyperref}       
\usepackage{booktabs}       
\usepackage{amsfonts}       
\usepackage{nicefrac}       
\usepackage{microtype}      
\usepackage{xcolor}         
\usepackage{url}
\usepackage{tabularx}
\usepackage{array}
\usepackage{footnote}
\usepackage{boldline}

\makesavenoteenv{tabular}
\makesavenoteenv{tabularx}

\hypersetup{
    colorlinks=true, 
    linkcolor=black, 
    citecolor=purple, 
    urlcolor=purple 
}

\title{A Review of the Evidence for Existential Risk from AI via Misaligned Power-Seeking}

\author{
Rose Hadshar \\
AI Impacts \\
\texttt{rosehadshar@gmail.com} 
}

\begin{document}
\maketitle

\begin{abstract}
Rapid advancements in artificial intelligence (AI) have sparked growing concerns among experts, policymakers, and world leaders regarding the potential for increasingly advanced AI systems to pose existential risks. This paper reviews the evidence for existential risks from AI via misalignment, where AI systems develop goals misaligned with human values, and power-seeking, where misaligned AIs actively seek power. The review examines empirical findings, conceptual arguments and expert opinion relating to specification gaming, goal misgeneralization, and power-seeking. The current state of the evidence is found to be concerning but inconclusive regarding the existence of extreme forms of misaligned power-seeking. Strong empirical evidence of specification gaming combined with strong conceptual evidence for power-seeking make it difficult to dismiss the possibility of existential risk from misaligned power-seeking. On the other hand, to date there are no public empirical examples of misaligned power-seeking in AI systems, and so arguments that future systems will pose an existential risk remain somewhat speculative. Given the current state of the evidence, it is hard to be extremely confident either that misaligned power-seeking poses a large existential risk, or that it poses no existential risk. The fact that we cannot confidently rule out existential risk from AI via misaligned power-seeking is cause for serious concern.

\end{abstract}


\newpage
\section{Executive summary}
Concerns that artificial intelligence could pose an existential risk are growing.

\textbf{This report reviews the evidence for existential risk from AI, focusing on arguments that future AI systems will pose an existential risk through misalignment and power-seeking:}

\begin{itemize}
\item \textit{Misalignment}: Some capable AI systems will develop goals which are misaligned with human goals.
\begin{itemize}
    \item \textit{Specification gaming}: Some capable AI systems will learn designer-specified goals which diverge from intended goals in unforeseen ways.
    \item \textit{Goal misgeneralization}: Some capable AI systems will develop goals which are perfectly correlated with intended goals in training, but diverge once the systems are deployed.
\end{itemize}
\item \textit{Power-seeking}: Some capable, misaligned AI systems will seek power in order to achieve their goals.
\end{itemize}
Our findings are based on a review of relevant literature, a series of interviews with AI researchers working on existential risk from AI \citep{noauthor_interviews_2023}, and a \href{https://wiki.aiimpacts.org/arguments_for_ai_risk/is_ai_an_existential_threat_to_humanity/database_of_empirical_evidence_about_ai_risk}{new database} of empirical evidence for some claims about existential risk from AI \citep{noauthor_empirical_nodate}.

\textbf{We find that the current state of the evidence for existential risk from misaligned power-seeking is concerning but inconclusive.}

\begin{itemize}
    \item There is strong empirical evidence of specification gaming and related phenomena, both in AI systems and other contexts, but it remains unclear whether specification gaming will be sufficiently extreme to pose an existential risk.
\item For goal misgeneralization, the evidence is more speculative. Examples of goal misgeneralization to date are sparse, open to interpretation, and not in themselves harmful. It’s unclear whether the evidence for goal misgeneralization is weak because it is not in fact a phenomenon which will affect AI systems, or because it will only affect AI systems once they are more goal-directed than at present.
\item There is also limited empirical evidence of power-seeking, but there are strong conceptual arguments and formal proofs which justify a stronger expectation that power-seeking will arise in some AI systems.

\end{itemize}

\textbf{Given the current state of the evidence, it is hard to be very confident either that misaligned power-seeking poses a large existential risk, or that it poses no existential risk.}

That we cannot confidently rule out existential risk from AI via misaligned power-seeking is cause for serious concern.

\newpage
\tableofcontents
\newpage

\section{Introduction}
\label{sec:Introduction}
Many claim that artificial intelligence could pose an existential risk - that \textbf{AI could lead to human extinction, or to a catastrophe which destroys humanity’s potential.}\footnote{Ord defines an existential catastrophe as “the destruction of humanity’s long-term potential” \citep{ord_precipice_2020}.}

Individual researchers have been making this claim for the last decade \citep{bostrom_superintelligence_2014,christian_alignment_2020,ord_precipice_2020,russell_human_2019}. More recently, the number of voices raising concerns about existential risk from AI has grown. In May 2023, hundreds of experts signed an open letter stating that “Mitigating the risk of extinction from AI should be a global priority alongside other societal-scale risks such as pandemics and nuclear war” \citep{noauthor_statement_2023}. Politicians have also begun to speak about the need to manage existential risk. For example, the UK’s Science, Innovation and Technology Committee has identified “the existential challenge” of AI as “a major threat to human life” as one of twelve areas for policymakers to address \citep{parliament2023}.

The argument that AI could pose an existential risk has been well made elsewhere \citep{bostrom_superintelligence_2014,carlsmith_is_2022,hendrycks_overview_2023,ord_precipice_2020}. The increasing prominence of the argument that AI could pose an existential risk, combined with the growing evidence base for some aspects of this argument, make now a good time to review the strength of the evidence for existential risk from AI.

\subsection{Scope}
\label{sec:Scope}
There are several different pathways to existential risk from AI.

The 2023 UK AI Safety Summit focuses on two of these pathways:\footnote{Some scholars have also pointed out a third pathway to existential risk from AI, via multi-agent interactions. See \cite{critch_ai_2020,drexler_reframing_2019,manheim_multiparty_2019}, and the \href{https://acsresearch.org/research}{Alignment of Complex Systems Research Group.}}
\begin{itemize}
    \item “\textbf{Misuse risks},\footnote{See \cite{hendrycks_overview_2023} for an introduction to misuse risks, which they term ‘Malicious use”.} for example where a bad actor is aided by new AI capabilities in biological or cyber-attacks, development of dangerous technologies, or critical system interference”
    \item “\textbf{Loss of control risks} that could emerge from advanced systems that we would seek to be aligned with our values and intentions” \citep{parliament2023}
\end{itemize}

A particular class of loss of control risks is \textbf{risks from misaligned power-seeking} \citep{carlsmith_is_2022}. The basic argument for existential risk from misaligned power-seeking is that:\footnote{See \hyperref[sec:A]{Appendix A} for a discussion of the more detailed argument given in \cite{carlsmith_is_2022}.}

\begin{itemize}
    \item (\textit{Preconditions}) In the not-too-distant future, some AI systems will be sufficiently capable to pose an existential risk.
    \item (\textit{Misalignment}) Some capable AI systems will develop goals which are misaligned with human goals.
\item (\textit{Power-seeking}) Some capable, misaligned AI systems will seek power in order to achieve their goals.
\item (\textit{Existential consequences}) This misaligned power-seeking will lead to human disempowerment, which will constitute an existential catastrophe.
\end{itemize}

\textbf{This report reviews the evidence for existential risk from future AI systems via misalignment and power-seeking.}

The following table breaks down the argument for existential risk from misaligned power-seeking further, and highlights the areas which are in the scope of this report.

\begin{table}[ht!]
\centering
\caption{The argument for existential risk from misaligned power-seeking}
\label{table:1}
\begin{tabularx}{\textwidth}{
   >{\raggedright\arraybackslash}X 
   >{\raggedright\arraybackslash}X }
 \hlineB{3}
 \textbf{Preconditions:} In the not-too-distant future, some AI systems will be sufficiently capable to pose an existential risk. & 
 \begin{itemize}
    \item \textit{Timelines:} The relevant AI systems will be developed in the not-too-distant future.
    \item \textit{Capabilities:} Some AI systems will be highly capable, in the sense that they are able to perform many important tasks at or above human level.
    \item \textit{Goal-directedness:} Some AI systems will be goal-directed, in that they pursue goals consistently over long time periods.
    \item \textit{Situational awareness:}\footnote{“A model is situationally aware if it's aware that it's a model and can recognize whether it's currently in testing or deployment.” \cite{berglund_taken_2023}} Some AI systems will be aware that they are AI systems, and whether they are in training or deployment.
 \end{itemize} \\
 \hline
 \textbf{Misalignment:}\footnote{“An AI is misaligned whenever it chooses behaviors based on a reward function that is different from the true welfare of relevant humans.” \cite{hadfield-menell_incomplete_2019}} Some capable AI systems will develop goals which are misaligned with human goals. & 
\begin{itemize}
    \item \textbf{Specification gaming:}\footnote{"Specification gaming is a behaviour that satisfies the literal specification of an objective without achieving the intended outcome." \cite{krakovna_specification_2020-1}. Specification gaming is related to proxy gaming \cite{hendrycks_overview_2023}, side effects \cite{amodei_concrete_2016,leike_ai_2017}, reward gaming \cite{leike_ai_2017}, reward hacking \cite{amodei_concrete_2016,skalse_defining_2022}, reward misspecification \cite{ngo_alignment_2023}, and Goodhart’s law \cite{hennessy_goodharts_2023,manheim_categorizing_2019,thomas_reliance_2022}.
} Some capable AI systems will learn designer-specified goals which diverge from intended goals in unforeseen ways.
    \item \textbf{Goal misgeneralization:}\footnote{"Goal misgeneralization is a specific form of robustness failure for learning algorithms in which the learned program competently pursues an undesired goal that leads to good performance in training situations but bad performance in novel test situations." \cite{shah_goal_2022}. Goal misgeneralization is related to goal drift \cite{hendrycks_overview_2023} and distributional shift \cite{amodei_concrete_2016,leike_ai_2017}.} Some capable AI systems will develop goals which are perfectly correlated with intended goals in training, but diverge once the systems are deployed.
\end{itemize}
\\ 
\hline
 \textbf{Power-seeking:}\footnote{Here we follow \cite{carlsmith_is_2022} and define power-seeking as “active efforts by an AI system to gain and maintain power in ways that designers didn’t intend, arising from problems with that system’s objectives."} Some capable, misaligned AI systems will seek power in order to achieve their goals. & \\ 
 \hline
 \textbf{Existential consequences:} This misaligned power-seeking will lead to human disempowerment, which will constitute an existential catastrophe. &
 \begin{itemize} 
    \item \textit{Disempowerment:} This misaligned power-seeking will lead to permanent human disempowerment.
    \item \textit{Existential catastrophe:} Permanent human disempowerment will constitute an existential catastrophe.
 \end{itemize} \\
 \hlineB{3}
\end{tabularx}
\end{table}

\hyperref[sec:B]{Appendix B} gives a shallow review of the evidence for some further claims about existential risk from AI which are outside of the scope of this report. 
\newpage
\subsection{Methodology}
\label{sec:Methodology}
This report is based on:
\begin{enumerate}
    \item \textbf{A review of the relevant literature on misaligned power-seeking}

    \item \textbf{A series of interviews with AI researchers working on existential risk from AI}

    We interviewed six AI researchers about the strength of the evidence for existential risk from AI. Summaries and recordings of some of the interviews can be found \href{https://wiki.aiimpacts.org/arguments_for_ai_risk/is_ai_an_existential_threat_to_humanity/interviews_on_the_strength_of_the_evidence_for_ai_risk_claims}{here}.

    Note that the sample size is small, and we did not interview AI researchers who are skeptical of existential risk from AI.\footnote{ We didn’t have the resources to interview a representative sample, and decided that we would get the most relevant information from speaking with researchers who work on AI existential risk and so are familiar with the evidence.}

    \item \textbf{A new database of empirical evidence for some claims about existential risk from AI}

    The full database can be accessed \href{https://wiki.aiimpacts.org/arguments_for_ai_risk/is_ai_an_existential_threat_to_humanity/database_of_empirical_evidence_about_ai_risk}{here}. It covers empirical evidence only, and includes evidence relating to specification gaming, goal misgeneralization and power-seeking (as well as deceptive alignment, self-improvement, and other claims relating to existential risk from AI).

    The database draws significantly from existing databases on \href{https://docs.google.com/spreadsheets/d/e/2PACX-1vRPiprOaC3HsCf5Tuum8bRfzYUiKLRqJmbOoC-32JorNdfyTiRRsR7Ea5eWtvsWzuxo8bjOxCG84dAg/pubhtml?urp=gmail_link&gxids=7628}{specification gaming} \citep{krakovna_specification_2020} and \href{https://docs.google.com/spreadsheets/d/e/2PACX-1vTo3RkXUAigb25nP7gjpcHriR6XdzA_L5loOcVFj_u7cRAZghWrYKH2L2nU4TA_Vr9KzBX5Bjpz9G_l/pubhtml}{goal misgeneralization} \citep{shah_goal_2022-1}.
\end{enumerate}
\newpage
\section{A review of the evidence for existential risk from misaligned power-seeking}
\label{sec:review}

Most of the AI existential risk researchers we interviewed regarded the evidence for misaligned power-seeking as at least somewhat speculative or uncertain.
\footnote{
“The main best objection I get from really smart people on this is that most of the evidence is of a weaker or more speculative form than what we are used to using to evaluate policies, at least really expensive policies like the ones AI doomers are advocating. They basically say, if I believed you based on these sorts of arguments, I would also have to believe lots of other people saying crazy sounding things. And I think they’re right that this is actually a weaker form of evidence that’s easier to spoof.” [36:07]  \citep{noauthor_interview_2023}

“I think that evidence for goal-directedness and correspondingly power-seeking is weaker. There’s kind of a cluster of arguments that are based on systems being goal-directed, both real goal misgeneralization and intentional power-seeking, and so on. And that's something that we're more uncertain about… deceptive alignment is also part of that cluster because that also relies on the system developing more goal-directedness.” [56:25] \citep{noauthor_interview_2023-2}

“The arguments about misalignment risk are definitely more uncertain in that they are doing more extrapolation. Both arguments are doing extrapolation. I think the misalignment stuff is sometimes doing a bit more of a difficult extrapolation, because it’s extrapolating these generalization properties which is just notoriously hard to do. I think that means that the case is just much more uncertain, but the case that the stakes are big is very good.” [47:16] \citep{noauthor_interview_2023-1}
}
Below, we review the evidence for misaligned power-seeking, including both conceptual and empirical evidence.

\subsection{The strength of the empirical evidence}
In general, the empirical evidence is weaker than the conceptual arguments for these claims about existential risk from AI. This is discussed in the relevant sections, but there are also some general points to make about the relative weakness of empirical evidence for misaligned power-seeking.
\label{sec:empirical}

Firstly, there are other properties of AI systems which might prove to be preconditions of misaligned power-seeking, but which current systems have not yet attained. It is plausible that systems will only display misaligned power-seeking at higher levels of general capabilities for example,\footnote{“The story of you train an AI to fetch a coffee and then it realizes that the only way it can do that is to take over the world is a story about misgeneralization. And it's happening at a very high level of abstraction. You're using this incredibly intelligent system which is reasoning at a very high level about things and it's making the error at that high level... And I think the state of the evidence is… we've never observed a misgeneralization failure at such a high level of abstraction, but that's what we would expect because we don't have AIs that can even reason at that kind of level of abstraction.” [28:36] \cite{noauthor_interview_2023-1}} or that misaligned power-seeking requires a higher level of goal-directedness than current systems have.\footnote{“What I'm expecting is happening here is that current systems are not goal-directed enough to show real power-seeking. And so the power-seeking threat model becomes more reliant on these kind of extrapolations of when there are systems which are more capable, they'll probably be at least somewhat more goal-directed and then once we have goal-directedness, we can more convincingly argue that power-seeking is going to be a thing because we have theory and so on, but there's a lot of uncertainty about it because we don't know how much systems will become more goal-directed.” [54:35] \citep{noauthor_interview_2023-2}}

Secondly, several of the AI researchers we interviewed clarified that the empirical evidence so far forms only a small or very small part of their reasons for concern about misaligned power-seeking, with more weight placed on conceptual arguments.\footnote{ “[Hadshar] Empirical details about capabilities that AI systems have now don’t sound very important to your world view.
[Researcher] Exactly.” [30:08] \citep{noauthor_interview_2023}

“I think that theoretical or conceptual arguments do have a lot of weight. Maybe I would put that at 60\% and empirical examples at 40\%, but I'm pulling this out of the air a little bit.” [24:00] \citep{noauthor_interview_2023-2}}

\subsection{The evidence for misalignment}
\label{sec:misalignment}
In this report, we consider two routes to capable AI systems developing goals which are misaligned with human goals:
\begin{itemize}
    \item \textbf{Specification gaming},\footnote{ "Specification gaming is a behavior that satisfies the literal specification of an objective without achieving the intended outcome." \citep{krakovna_specification_2020-1}. Specification gaming is related to proxy gaming \citep{hendrycks_overview_2023}, side effects \citep{amodei_concrete_2016,leike_ai_2017}, reward gaming \citep{leike_ai_2017}, reward hacking \citep{amodei_concrete_2016,skalse_defining_2022}, reward misspecification \citep{ngo_alignment_2023}, and Goodhart’s law \citep{hennessy_goodharts_2023,manheim_categorizing_2019,thomas_reliance_2022}.} where some capable AI systems learn designer-specified goals which diverge from intended goals in unforeseen ways.
    \item \textbf{Goal misgeneralization},\footnote{"Goal misgeneralization is a specific form of robustness failure for learning algorithms in which the learned program competently pursues an undesired goal that leads to good performance in training situations but bad performance in novel test situations." \citep{shah_goal_2022}. Goal misgeneralization is related to goal drift \citep{hendrycks_overview_2023} and distributional shift \citep{amodei_concrete_2016,leike_ai_2017}.
} where some capable AI systems develop goals which are perfectly correlated with intended goals in training, but diverge once the systems are deployed.
\end{itemize}

\subsubsection{The evidence for specification gaming}
\label{sec:specification}
One route to AI systems developing misaligned goals is specification gaming, where AI systems learn the goals which they are given, but these goals are misspecified and come apart from intended goals.

"Specification gaming is a behavior that satisfies the literal specification of an objective without achieving the intended outcome." \citep{krakovna_specification_2020-1} If sufficiently powerful AI systems were to be deployed in high-stakes settings, then the difference between the literal specification and the intended outcome could become extreme, leading to catastrophic outcomes \citep{bostrom_superintelligence_2014,pueyo_growth_2018}.

Specification gaming is a well-established phenomenon, both in general and in the context of AI systems. 

In non-AI contexts, there are numerous examples of variants of specification gaming,\footnote{ For discussions about a cluster of related concepts including Goodhart’s Law and proxy failure, see \cite{amodei_concrete_2016,john_dead_2023,manheim_categorizing_2019,thomas_reliance_2022}.} in economics \citep{braganza_proxyeconomics_2022,chrystal_goodharts_2003,goodhart_problems_1984,kelly_capitalism_2021,lucas_econometric_1976}, education \citep{berliner_inevitable_2005,campbell_assessing_1979,elton_goodharts_2004,fire_over-optimization_2019,koretz_measuring_2008,strathern_improving_1997,stroebe_why_2016}, healthcare \citep{omahony_medicine_2017,poku_campbells_2016} and other areas.\footnote{ See Table 1 in \cite{john_dead_2023} for a collection of examples.} It is clear that at least in human and social systems, such dynamics are widespread.

In the context of AI systems, there are both theoretical demonstrations of specification gaming given certain model assumptions \citep{beale_unethical_2020,hennessy_goodharts_2023,manheim_categorizing_2019,zhuang_consequences_2021}, and many empirical examples of specification gaming in AI systems, both in toy environments and in deployment \citep{krakovna_specification_2020-1}.\footnote{The \href{https://docs.google.com/spreadsheets/d/e/2PACX-1vRPiprOaC3HsCf5Tuum8bRfzYUiKLRqJmbOoC-32JorNdfyTiRRsR7Ea5eWtvsWzuxo8bjOxCG84dAg/pubhtml}{database} linked to from this post contains over 70 examples of specification gaming. See also \cite{noauthor_empirical_nodate}.

“One form of the problem has also been studied in the context of feedback loops in machine learning systems (particularly ad placement), based on counterfactual learning and contextual bandits. The proliferation of reward hacking instances across so many different domains suggests that reward hacking may be a deep and general problem, and one that we believe is likely to become more common as agents and environments increase in complexity.” \citep{amodei_concrete_2016}.
“Reward hacking—where RL agents exploit gaps in misspecified reward functions—has been widely observed” \citep{pan_effects_2022}.
}

For example, OpenAI trained an agent to play the game CoastRunners. The agent was rewarded for hitting targets along the course of a boat race. But instead of racing to the finish line, the agent discovered a loophole where it could race in a circle, repeatedly crashing and setting itself on fire, to earn maximum points \citep{noauthor_faulty_2016}.

While a majority of clear examples of specification gaming in AI systems arise in toy environments like CoastRunners \citep{krakovna_specification_2020}, there are already some examples of deployed AI systems engaging in specification gaming, and of this behavior leading to harm, particularly in the areas of bias and misinformation. 

For example, a healthcare screening system deployed in 2019 was trained to predict health care costs. As less is spent on Black patients’ care because of unequal access to healthcare, the algorithm rated Black patients as less sick than White patients even where Black patients had more underlying chronic illnesses \citep{obermeyer_dissecting_2019}. 

Falsehoods generated by large language models can also be viewed as the result of specification gaming, though here the case is less clear. Language models trained to accurately predict the next token frequently generate false content \citep{noauthor_incident_2016,noauthor_incident_2016-1,noauthor_why_2022}, but as one of our interviewees pointed out, it is a matter of judgment whether this is best interpreted as specification gaming or as a simple capability failure.\footnote{“With some of the language model examples, I think you can ask the question, is this really specification gaming, or is it capability failure, or something like that? I think sometimes there's a bit of a judgment call there.” [29:45] \citep{noauthor_interview_2023-2}}

The evidence is strong that AI systems will be subject to specification gaming to some degree. It remains unclear whether specification gaming will be sufficiently serious to pose an existential risk. In order to cause large-scale harms, misspecified goals would need to be subtle enough that systems were still deployed in high-stakes settings, but diverge extremely from intended goals in deployment. To date, no examples of specification gaming in AI systems have been catastrophic, so there is no direct evidence of this degree of harm from specification gaming.

There are some tentative signs that specification might become a more serious problem as models become more capable. In initial experiments, larger language models and language models with more RLHF are more prone to sycophantic answers, and to expressing a desire to seek power and avoid shutdown \citep{perez_discovering_2022}. Insofar as these behaviors are indeed caused by specification gaming,\footnote{That is, the systems are following the specified goal of generating text which receives high positive feedback from humans, but this comes apart from the goal of generating helpful, honest and harmless text. See also \cite{krakovna_specification_2020}.} this is cause for concern. Another study has found that when goals are misspecified, more capable RL agents will diverge more from intended goals than less capable agents, suggesting that specification gaming may worsen as capabilities improve. The same study also found that the divergence between intended and misspecified goals was sometimes very sudden, which might make it hard to anticipate and prevent such problems arising in deployment \citep{pan_effects_2022}. 

Overall, the evidence for specification gaming is strong, though it remains unclear whether the scale of the problem will be sufficient to pose an existential risk.

\subsubsection{The evidence for goal misgeneralization}
Another route to AI systems developing misaligned goals is goal misgeneralization, where systems develop goals which are perfectly correlated with intended goals in training, but diverge once the systems are deployed.

"Goal misgeneralization is a specific form of robustness failure for learning algorithms in which the learned program competently pursues an undesired goal that leads to good performance in training situations but bad performance in novel test situations." \citep{shah_goal_2022}

The underlying mechanism behind goal misgeneralization is distributional shift, where there are systematic differences between the training distribution and the test distribution. Distributional shift is a very widely documented phenomenon in AI systems \citep{leike_ai_2017,quinonero-candela_dataset_2022}, and out-of-distribution robustness remains unsolved \citep{hendrycks_many_2021,liu_towards_2023}. This provides a reason to expect goal misgeneralization to arise.

However, the empirical evidence for goal misgeneralization is currently weak, in spite of the prevalence of distributional shift.

There are examples of goal misgeneralization in AI systems \citep{degrave_ai_2021,langosco_goal_2023,shah_goal_2022}. However, these examples do not conclusively show that goal misgeneralization will arise in a harmful way.

Firstly, all of the examples of goal misgeneralization we have found take place in demonstration, rather than in deployed systems. Sometimes these demonstrations involve very obvious and crude differences between the training data and the test data. For instance, \cite{langosco_goal_2023} train a CoinRun agent exclusively on mazes where the cheese is always in the upper right hand corner, and show in testing that the agent learns to navigate to the upper right rather than to the cheese. This shows that goal misgeneralization can occur when the training data is very different to the test data - but doesn’t provide evidence for goal misgeneralization in more realistic settings. We have not found any evidence of real-world harm from goal misgeneralization so far.

Secondly, it is currently not possible to demonstrate conclusively that examples of goal misgeneralization actually involve systems learning a goal which is correlated in training but not deployment. It is only possible to observe the behavior of the system in question, not its inner workings, so we cannot know what goal (if any) a system has learned. Examples to date only conclusively show behavioral or functional goal misgeneralization.\footnote{ “I think right now the examples we have are more like behavioral goal misgeneralization where you just have different behaviors that are all the same in training but then they become decoupled in the new setting but we don't know how the behavior is going to generalize. We call it goal misgeneralization maybe more as a shorthand. The behavior has different ways of generalizing that are kind of coherent. We can present it as the system learned the wrong goal, but we can't actually say that it has learned a goal. Maybe it’s just following the wrong heuristic or something. I think the current examples are a demonstration of the more obvious kind of effect where the training data doesn't distinguish between all the ways that the behavior could generalize.” [37:11] \citep{noauthor_interview_2023-2}
}

Furthermore, it’s often hard to distinguish goal misgeneralization from capability misgeneralization, where the system’s capabilities also fail to generalize.\footnote{“I think it's a less well understood phenomenon… it can be hard to distinguish capability misgeneralization from goal misgeneralization.” [33:16] \citep{noauthor_interview_2023-2}
} In the abstract, goal misgeneralization is distinct from capability misgeneralization: “a system’s capabilities generalize but its goal does not generalize as desired. When this happens, the system competently pursues the wrong goal.” \citep{shah_goal_2023} But in real-world settings, the wrong goal may often lead to capability failure. A system which learns to competently predict that tumors with rulers are malignant based on its training data will fail to competently predict actual malignancy when tested on more diverse data \citep{narla_automated_2018}. Insofar as goal misgeneralization comes with capability misgeneralization, AI systems which learn very misgeneralized goals are unlikely to be deployed.

There are several possible explanations of the weakness of evidence on goal misgeneralization so far.

Goal misgeneralization might require a level of goal-directedness which current systems don’t yet have,\footnote{“Specifying something as goal misgeneralization also requires some assumption that the system is goal-directed to some degree and that can also be debatable.” [33:16] \citep{noauthor_interview_2023-2}} or an ability to reason at higher levels of abstraction.\footnote{“The story of you train an AI to fetch a coffee and then it realizes that the only way it can do that is to take over the world is a story about misgeneralization. And it's happening at a very high level of abstraction. You're using this incredibly intelligent system which is reasoning at a very high level about things and it's making the error at that high level... And I think the state of the evidence is… we've never observed a misgeneralization failure at such a high level of abstraction, but that's what we would expect because we don't have AIs that can even reason at that kind of level of abstraction.” [28:36] \citep{noauthor_interview_2023-1}} Reliably identifying goal misgeneralization might also require more advanced interpretability techniques.\footnote{“The mechanism is a lot less well understood. I think to really properly diagnose goal misgeneralization we would need better interpretability tools.” [36:30] \citep{noauthor_interview_2023-2}} Alternatively, the distinction between behavioral and ‘actual’ goal misgeneralization may be misplaced: if sufficiently capable systems engage in behaviors which look like goal misgeneralization, then functionally they are misaligned whether or not their internal representations match our description of goal misgeneralization.

So there are some reasons to expect the current evidence of goal misgeneralization to be weak, even if the phenomenon eventually arises strongly. Nevertheless, so far the evidence for goal misgeneralization remains reasonably speculative.\footnote{“I think [the evidence for goal misgeneralization] is not as strong [as for specification gaming].” [33:16]  \citep{noauthor_interview_2023-2}
 “These generalization failures at new levels of abstraction are notoriously hard to predict. You have to try and intuit what an extremely large scale neural net will learn from the training data and in which ways it will generalize… I’m relatively persuaded that misgeneralization will continue to happen at higher levels of abstraction, but whether that actually is well described by some of the typical power-seeking stories I’m much less confident and it’s definitely going to be a judgment call.” [28:36] \citep{noauthor_interview_2023-1}
}

\subsection{The evidence for power-seeking}

The presence of misaligned goals in and of itself need not pose an existential risk. But if AI systems with misaligned goals successfully and systematically seek power, the result could be existential.

In \cite{carlsmith_is_2022}, power-seeking is defined as “active efforts by an AI system to gain and maintain power in ways that designers didn’t intend, arising from problems with that system’s objectives." 

Carlsmith loosely defines power as “the type of thing that helps a wide variety of agents pursue a wide variety of objectives in a given environment.” \citep{carlsmith_is_2022} We can take Bostrom’s categories of instrumental goals as illustrative of this “type of thing”:

\begin{itemize}
    \item Self-preservation
    \item Goal-content integrity\footnote{“An agent is more likely to act in the future to maximize the realization of its present final goals if it still has those goals in the future. This gives the agent a present instrumental reason to prevent alterations of its final goals.” \citep{bostrom_superintelligent_2012}}
    \item Cognitive enhancement
    \item Technological perfection\footnote{“An agent may often have instrumental reasons to seek better technology, which at its simplest
means seeking more efficient ways of transforming some given set of inputs into valued outputs.” \citep{bostrom_superintelligent_2012}}
    \item Resource acquisition \citep{bostrom_superintelligent_2012}
\end{itemize}

\textbf{The conceptual argument that some AI systems will seek power seems strong.}\footnote{“I think some of the other theoretical arguments like instrumental convergence also generally seems like a very clear argument, and we can observe some of these effects in human systems and corporations and so on.” [25:23] \citep{noauthor_interview_2023-2}} Bostrom’s instrumental convergence thesis is simple and intuitively plausible: “as long as they possess a sufficient level of intelligence, agents having any of a wide range of final goals will pursue similar intermediary goals because they have instrumental reasons to do so.” \citep{bostrom_superintelligent_2012}

There are formal proofs that the instrumental convergence thesis holds for various kinds of AI systems. \cite{turner_optimal_2023} prove that “most reward functions make it optimal to seek power by keeping a range of options available” in the context of Markov decision processes. \cite{turner_parametrically_2022} extend this result to a class of sub-optimal policies, showing that “many decision-making functions are retargetable, and that retargetability is sufficient to cause power-seeking tendencies”. \cite{krakovna_power-seeking_2023} further show that agents which learn a goal are likely to engage in power-seeking.

The formal and theoretical case for power-seeking in sufficiently capable and goal-directed AI systems is therefore relatively strong.

\textbf{However, the empirical evidence of power-seeking in AI systems is currently weak}. There are some demonstrations of RL agents engaging in power-seeking behaviors in toy environments (for example, \cite{hadfield-menell_off-switch_2017}), but no convincing examples of AI systems in the real world seeking power in this way to date.\footnote{“I don’t think there’s really empirical evidence [for power-seeking]... To me it’s very uncertain.” [28:36] \citep{noauthor_interview_2023-1}}

\cite{perez_discovering_2022} show language models giving “answers that indicate a willingness to pursue potentially dangerous subgoals: resource acquisition, optionality preservation, goal preservation, powerseeking, and more." But indicating willingness is not the same as actually engaging in power-seeking behaviors. Language models might express power-seeking desires merely because their training data contains similar text, and not because they will ever directly seek power.

Sycophancy, where language models agree with their users regardless of the accuracy of the statements, could be taken as an example of power-seeking behavior. But as with the results of \cite{perez_discovering_2022}, sycophancy is likely to be simply an imitation of the training data, rather than an intentional behavior.\footnote{“Looking at current systems, sycophancy can be considered as a form of power-seeking. Although I think that's also maybe debatable. It's building more influence with the user by agreeing with their views, but it's probably more of a heuristic that is just somehow reinforced than intentional power-seeking.” [49:35] \citep{noauthor_interview_2023-2}
}

If the theoretical arguments for power-seeking are strong, why is the empirical evidence to date weak?

As with goal misgeneralization, one plausible explanation is that power-seeking behavior depends on a level of goal-directedness or capability in general which current models don’t yet have.\footnote{“What I'm expecting is happening here is that current systems are not goal-directed enough to show real power-seeking. And so the power-seeking threat model becomes more reliant on these kind of extrapolations of when there are systems which are more capable, they'll probably be at least somewhat more goal-directed and then once we have goal-directedness, we can more convincingly argue that power-seeking is going to be a thing because we have theory and so on, but there's a lot of uncertainty about it because we don't know how much systems will become more goal-directed.” [54:35] \citep{noauthor_interview_2023-2}}

Overall, with strong conceptual arguments but no public empirical evidence, it seems plausible but unproven that some AI systems will seek power.

\newpage
\section{Conclusion: The current strength of the evidence for existential risk from misaligned power-seeking}

The current state of the evidence for existential risk from misaligned power-seeking is concerning but inconclusive.

There is strong empirical evidence of specification gaming and related phenomena, both in AI systems and other contexts. We can be reasonably confident therefore that specification gaming will arise to some extent in future AI systems, but it remains unclear whether specification gaming will be sufficiently extreme to pose an existential risk.

For goal misgeneralization, the evidence is more speculative. Distributional shift, which is a prerequisite of goal misgeneralization, is a well-documented phenomenon, but the examples of goal misgeneralization to date are sparse, open to interpretation, and not in themselves harmful. It’s unclear whether there is weak evidence for goal misgeneralization because it is not in fact a phenomenon which will affect AI systems to a harmful degree, or because it will only affect AI systems once they are more goal-directed than at present.

There is also limited empirical evidence of power-seeking, but there are strong conceptual arguments and formal proofs which justify a stronger expectation that power-seeking will arise in some AI systems.

Strong empirical evidence of specification gaming combined with strong conceptual arguments for power-seeking make it difficult to dismiss the possibility of existential risk from misaligned power-seeking. On the other hand, we are not aware of any empirical examples of misaligned power-seeking in AI systems, and so arguments that future systems will pose an existential risk must remain somewhat speculative.

Given the current state of the evidence, it is hard to be extremely confident either that misaligned power-seeking poses a large existential risk, or that it poses no existential risk.

That we cannot confidently rule out existential risk from AI via misaligned power-seeking is cause for serious concern.

\newpage
\section{Acknowledgements}

Thanks to Katja Grace and Harlan Stewart in particular; to Michael Aird, Adam Bales, Rick Korzekwa, Fazl Barez, Sam Clark, Max Dalton, and many others for various levels of feedback and support; and to all the researchers we interviewed.

\section{References}
\begingroup
\renewcommand
\refname{}
\bibliography{references}
\bibliographystyle{acl_natbib}  
\endgroup

\newpage
\section{Appendix A: Carlsmith’s argument for existential risk via power-seeking AI}
\label{sec:A}
The following table maps between the premises of \cite{carlsmith_is_2022}’s argument, and the claims used in this report (see Table 1). Claims within the scope of this report are bolded. 

Note that the claims used in this report are not identical to Carlsmith’s premises, though they are closely related.

\begin{longtable}{p{0.175\textwidth}|p{0.375\textwidth}|p{0.35\textwidth}}
    \hlineB{3}  
    \multicolumn{2}{l}{\textbf{Carlsmith}} & {\textbf{Claims used in this report}} \\
    \hline
    \multicolumn{2}{l}{By 2070:} & (\textit{Preconditions: Timelines}) The relevant AI systems will be developed in the not-too-distant future. \\
    \hline
    1. It will become possible and financially feasible to build AI systems with the following properties: &
    \textit{Advanced capability}: they outperform the best humans on some set of tasks which when performed at advanced levels grant significant power in today’s world (tasks like scientific research, business/military/political strategy, engineering, and persuasion/manipulation). &
    \textit{(Preconditions: Capabilities}) Some AI systems will be highly capable, in the sense that they are able to perform many important tasks at or above human level \\
    & \newline \textit{Agentic planning}: they make and execute plans, in pursuit of objectives, on the basis of models of the world. &  \newline (\textit{Preconditions: Goal-directedness}) Some AI systems will be goal-directed, in that they pursue goals consistently over long time periods. \\

    & \newline \textit{Strategic awareness}: the models they use in making plans represent with reasonable accuracy the causal upshot of gaining and maintaining power over humans and the real-world environment. \newline \newline (Call these “APS”—Advanced, Planning, Strategically aware—systems.) & \newline (\textit{Preconditions: Situational awareness}) Some AI systems will be aware that they are AI systems, and whether they are in training or deployment. \\

    \hline

    \multicolumn{2}{p{0.6\textwidth}}{2. There will be strong incentives to build and deploy APS systems.} & \\
    \hline
    \multicolumn{2}{p{0.6\textwidth}}{3. It will be much harder to build APS systems that would not seek to gain and maintain power in unintended ways (because of problems with their objectives) on any of the inputs they’d encounter if deployed, than to build APS systems that would do this, but which are at least superficially attractive to deploy anyway.} &
    (\textit{Misalignment}) Some capable AI systems will develop goals which are misaligned with human goals.\newline \newline (\textit{Misalignment: Specification gaming}) Some capable AI systems will learn designer-specified goals which diverge from intended goals in unforeseen ways. \newline \newline (\textit{Misalignment: Goal misgeneralization}) Some capable AI systems will develop goals which are perfectly correlated with intended goals in training, but diverge once the systems are deployed.
    \\
    \hline
    \multicolumn{2}{p{0.6\textwidth}}{4. Some deployed APS systems will be exposed to inputs where they seek power in unintended and high-impact ways (say, collectively causing >\$1 trillion dollars of damage), because of problems with their objectives.} & (\textit{Power-seeking}) Some capable, misaligned AI systems will seek power in order to achieve their goals.\\
    \hline
    \multicolumn{2}{p{0.6\textwidth}}{5. Some of this power-seeking will scale (in aggregate) to the point of permanently disempowering ~all of humanity.} & (\textit{Existential consequences: Disempowerment}) This misaligned power-seeking will lead to permanent human disempowerment.\\
    \hline
    \multicolumn{2}{p{0.6\textwidth}}{6. This disempowerment will constitute an existential catastrophe.} & (\textit{Existential consequences: Existential catastrophe}) Permanent human disempowerment will constitute an existential catastrophe.\\
    \hlineB{3}  
\end{longtable}

\newpage
\section{Appendix B: Some evidence for other claims about existential risk from AI}
\label{sec:B}

We systematically reviewed the evidence for claims about misalignment and power-seeking. However, in the course of our research and interviews, we came across some evidence for other relevant claims.

This appendix contains some of the evidence for goal-directedness, situational awareness, and deceptive alignment. It should not be treated as a comprehensive review of the state of the evidence on these topics.

\subsection{Some evidence for goal-directedness}

Roughly, goal-directedness refers to a property of AI systems to persistently pursue a goal.
\footnote{In \cite{carlsmith_is_2022}, goal-directedness is referred to as “agentic planning”, where AI systems “make and execute plans, in pursuit of objectives, on the basis of models of the world.”}
Goal-directedness has not been well-defined so far, and so reviewing the evidence for goal-directedness is hampered by unclarity about the concept.
\footnote{ “Right now it's really hard to distinguish between real goal-directedness and learned heuristics… I think part of the problem with goal-directedness is we don’t really understand the phenomenon that well.” [44:00] \citep{noauthor_interview_2023-2}}

That said, it seems plausible that goal-directedness is a direct precondition for goal misgeneralization and for power-seeking,\footnote{“Specifying something as goal misgeneralization also requires some assumption that the system is goal-directed to some degree and that can also be debatable.” [33:16] \citep{noauthor_interview_2023-2}
“What I'm expecting is happening here is that current systems are not goal-directed enough to show real power-seeking. And so the power-seeking threat model becomes more reliant on these kind of extrapolations of when there are systems which are more capable, they'll probably be at least somewhat more goal-directed and then once we have goal-directedness, we can more convincingly argue that power-seeking is going to be a thing because we have theory and so on, but there's a lot of uncertainty about it because we don't know how much systems will become more goal-directed.” [54:35] \citep{noauthor_interview_2023-2}} so it is an important claim to assess.

Coherence theorems offer one kind of conceptual evidence for goal-directedness, but the extent to which they apply to future AI systems is contested \citep{bales_will_2023,ejt_there_2023,noauthor_what_2021}.\footnote{“Some of the theoretical arguments make the case that goal-directedness is an attractor. I think that's something that's more debatable, less clear to me. There have been various discussions on LessWrong and elsewhere about to what extent do coherence arguments imply goal-directedness. And I think the jury is still out on that one.” [42:36] \citep{noauthor_interview_2023-2}}

There is limited empirical evidence of goal-directedness in systems so far.\footnote{“I think the evidence so far at least for language models, there isn't really convincing evidence of goal-directedness.” [44:00] \citep{noauthor_interview_2023-2}} One of the researchers we interviewed noted that language models may be particularly unsuited to goal-directedness.\footnote{ “It’s also possible goal-directedness is kind of hard. And especially, maybe language models are just a kind of system where goal-directedness comes less naturally than other systems like reinforcement learning systems or even with humans or whatever.” [40:26] \citep{noauthor_interview_2023-2}}

However, individual researchers we interviewed believe that:

\begin{itemize}
    \item To the extent that language models can simulate humans, they will have the ability to simulate goal-directedness.\footnote{ “I think generally the kind of risk scenarios that we are most worried about would involve the system acting intentionally and deliberately towards some objectives but I would expect that intent and goal-directedness comes in degrees and if we see examples of increasing degrees of that then I think that does constitute evidence of that being possible. Although it’s not clear whether it will go all the way to really deliberate systems, but I think especially to the extent that these systems can simulate humans… they have the ability to simulate deliberate intentional action and planning because that's something that humans can do.” [20:20] \citep{noauthor_interview_2023-2}}
    \item There is a clear trend towards systems acting more autonomously.\footnote{ “We are already capable of getting AI systems to do simple things relatively autonomously. I don’t think it’s a threshold where now it’s autonomous, now it’s not… I think it’s a spectrum and it’s just very clearly ramping up. We already have things that have a little autonomy but not very much. I think it's just a pretty straightforward trend at this point.” [24:39] \citep{noauthor_interview_2023-1}} 
\end{itemize}

One researcher we interviewed highlighted goal-directedness as one of their key uncertainties about existential risk from AI.\footnote{ “I think we might see more goal-directed systems which produce clearer examples of internal goal misgeneralization, but also I wouldn't be that surprised if we don't see that. I think that's one of the big uncertainties I have about level of risk. How much can we expect goal-directedness to emerge?” [40:26] \citep{noauthor_interview_2023-2}}

\subsection{Some evidence for situational awareness}

“A model is situationally aware if it's aware that it's a model and can recognize whether it's currently in testing or deployment.” \citep{berglund_taken_2023}

This is important to arguments about existential risk from AI as situational awareness is plausibly a precondition for successful misaligned power-seeking: a model may need to understand its own situation at a sophisticated level in order to make plans which successfully disempower humans. In particular, situational awareness seems like a precondition for deceptive alignment.

There is some empirical work demonstrating situational awareness in large language models, but the results are inconclusive \citep{berglund_taken_2023,ngo_alignment_2023,perez_discovering_2022}. \cite{berglund_taken_2023} find that language models can perform out-of-context reasoning tasks, but only with particular training set ups and data augmentation. \cite{perez_discovering_2022} run various experiments to test awareness, and find that “the models we evaluate are not aware of at least some basic details regarding themselves or their training procedures.” On the other hand, \cite{langosco_goal_2023} use the same questions as \cite{perez_discovering_2022} but find that their model answers 85\% accurately.

\end{document}